\documentclass[preprint,showkeys,secnumarabic,amsfonts,amsmath,amssymb]{revtex4}
\usepackage[dvips]{color}
\usepackage{array}
\usepackage{epsfig}
\usepackage{amsmath}
\usepackage{graphicx}

\begin{document}

\title{The Stability Analysis of Brane Induced Gravity with Quintessence Field on the Brane with a Gaussian Potential}

\author{A. Ravanpak}
\email{a.ravanpak@vru.ac.ir}
\affiliation{Department of Physics, Vali-e-Asr University of Rafsanjan, Rafsanjan, Iran}

\author{G. F. Fadakar}
\email{g.farpour@vru.ac.ir}
\affiliation{Department of Physics, Vali-e-Asr University of Rafsanjan, Rafsanjan, Iran}

\date{\today}

\begin{abstract}

In this manuscript we consider a normal branch of the DGP cosmological model with a quintessence scalar field on the brane as the dark energy component. Using the dynamical system approach we study the stability properties of the model. We find that $\lambda$, as one of our new dimensionless variables which is defined in terms of the quintessence potential has a crucial role in the history of the universe. We divide our discussion into two parts: a constant $\lambda$, and a varying $\lambda$. In the case of a constant $\lambda$ we calculate all the critical points of the model even those at infinity and then assume all of them as instantaneous critical points in the varying $\lambda$ situation which is the main part of this manuscript. We find that the effect of the extra dimension in such a model is independent of the value of $\lambda$. Then, we consider a Gaussian potential for which $\lambda$ is not constant but varies from zero to infinity. We discuss the evolution of the dynamical variables of the model and conclude that their asymptotic behaviors follow the trajectories of the moving critical points. Also, we find two different possible fates for the universe. In one of them it could experience an accelerated expansion, but then enters a decelerating phase and finally reaches a stable matter dominated solution. In the other scenario, the universe could approach the matter dominated critical point without experiencing any accelerated expansion. We argue that the first scenario is more compatible with observations.

\end{abstract}


\maketitle

\section{Introduction}\label{sec1}

The outcomes of cosmological observations, such as the type Ia supernova (SNe Ia) \cite{Riess}, the cosmic microwave background radiation (CMBR) \cite{Spergel}, the large-scale structure \cite{Tegmark}, and so forth, have disclosed that our universe is undergoing an accelerated expanding phase. Cosmologists describe this surprising phenomenon either using the concept of dark energy (DE) \cite{Sahni}-\cite{Gao}, or with some extended theories of gravity \cite{Capozziello}-\cite{Bengochea}.

On the other hand, the concept of extra dimension that arises from string theory has attracted a great amount of attention, especially for explaining the so called hierarchy problem \cite{Hamed}-\cite{Saridakis}. In these theories our four dimensional (4D) universe is considered as a brane embedded in a five dimensional (5D) spacetime dubbed bulk. If a 4D scalar curvature term is added into the brane action, on top of its matter Lagrangian term, we are dealing with a brane induced gravity theory, and the work by Dvali, Gabadadze and Porrati called DGP braneworld model is its most well-known example in which the bulk is an infinite 5D Minkowski spacetime \cite{Dvali}. The DGP model includes two distinct branches, the self-accelerating branch that yields late-time acceleration geometrically, but suffers from the ghost instability, and the normal branch which is healthy, but cannot explain accelerated expansion of the universe without any DE component.

Furthermore, dynamical system analysis which could qualitatively be a practical method in examining the long-term behavior of the universe has been widely used in literature \cite{Wainwright}-\cite{Quiros2}. This qualitative study is based on the stability analysis. In this approach, instead of a particular trajectory, one finds various kinds of the possible trajectories of the universe in an appropriate phase space and categorizes them.

In this manuscript we will consider a normal branch of the DGP braneworld cosmology with a quintessence scalar field $\phi$ on the brane, as the DE component. To investigate this model, we will follow the dynamical system approach. After introducing some new dimensionless variables, we will write an autonomous system of ordinary differential equations. Then, we will obtain the critical points of the model and the related eigenvalues to study their stability. Although a dynamical investigation of the DGP model with a scalar field trapped on the brane has already been studied in \cite{Quiros2} for a constant and an exponential scalar field potential distinctly, but the prior is very simple and special, and the latter does not represent the effect of the extra dimension. Also, the critical points at infinity have not been studied in \cite{Quiros2}. Here, we will consider a Gaussian potential and show that with this selection of potential functions, not only our model can demonstrate different cosmological epochs such as the matter dominated phase and the DE dominated era, but also it can always represent the role of the extra dimension. Moreover, we will find a few moving critical points in our model that play an important role in the evolution of the universe.

This article is organized as follows: in Sec.\ref{sec2}, we review the basic equations of the model. Sec.\ref{sec3}, deals with the new variables and their respective autonomous differential equations. The critical points and their stability conditions are also discussed in this section, but for two different situations. In the first part of Sec.\ref{sec3}, we discuss a constant $\lambda$ case, and in its second part the case of a varying $\lambda$ is studied. The asymptotic behavior of the model is also investigated in this subsection. In the third part of Sec.\ref{sec3}, we try to compare the predictions of our model with observational data. Finally in Sec.\ref{sec4}, we express a summary and discuss the results.

\section{The Model}\label{sec2}

Assuming a homogeneous, isotropic and spatially flat brane in a normal branch of DGP model one can obtain the Friedmann equation on the brane as \cite{Deffayet}
\begin{equation}\label{fried}
H^2+\frac{H}{r_c}=\frac{1}{3M_p^2}(\rho_m+\rho_\phi)
\end{equation}
in which $\rho_m$, represents the energy density of the matter content of the universe, and $\rho_\phi$ demonstrates the enrgy density of the quintessence scalar field as the DE component. Also, $H$ and $M_p$ are the Hubble parameter and the Planck mass, respectively and $r_c$, is the crossover distance that determines the transition from 4D to 5D regime and is always positive. In the absence of interaction between the dark sectors of the universe, one can utilize the standard conservation equations as
\begin{eqnarray}
\dot\rho_m&+&3H\rho_m=0, \label{conservation1} \\ \dot\rho_\phi&+&3H(\rho_\phi+p_\phi)=0, \label{conservation2}
\end{eqnarray}
Here, the dot denotes derivative with respect to the cosmic time, $t$. In Eq.(\ref{conservation2}), the energy density of the quintessence scalar field $\rho_\phi$, and its pressure $p_\phi$, are defined as
\begin{eqnarray}
\rho_\phi &=& \frac{1}{2}\dot \phi^2+ V(\phi), \label{rp1} \\ p_\phi &=& \frac{1}{2}\dot \phi^2- V(\phi), \label{rp2}
\end{eqnarray}
respectively, in which $V(\phi)$ is the quintessence potential. Substituting $\rho_{\phi}$ and $p_{\phi}$, in Eq.(\ref{conservation2}), we obtain the equation of motion of the quintessence scalar field as
\begin{equation}\label{eom}
\ddot\phi+3H\dot\phi+V_{\phi}=0
\end{equation}
Here, the derivative of $V(\phi)$ with respect to the scalar field has been denoted by $V_{\phi}$.

\section{THE PHASE SPACE AND the STABILITY ANALYSIS}\label{sec3}

In order to analyze the stability characteristics of the model, we first introduce a set of new dimensionless variables to convert the equations of motion of our model into a self-autonomous dynamical system. The auxiliary variables we have selected are as follows:
\begin{eqnarray}\label{nv}
 x&=&\sqrt{\frac{\rho_m}{3M_p^2(H^2+\frac{H}{r_c})}}, \quad y=\sqrt{\frac{V(\phi)}{3M_p^2(H^2+\frac{H}{r_c})}}, \quad \\\nonumber z&=&\frac{\dot\phi}{\sqrt{6M_p^2(H^2+\frac{H}{r_c})}}, \quad l=\sqrt{1+\frac{1}{Hr_c}}, \quad \lambda=M_p V_\phi/V.
\end{eqnarray}
Since an expanding universe and a contracting one are independent submanifolds, we can study them separately \cite{Quiros2},\cite{Ravanpak2}. In the following, we will focus on the more popular case: a universe that is expanding. With $H>0$ for an expanding universe, and $r_c>0$, we find a constraint as $l\geq1$. One can check that for $r_c \rightarrow \infty$, we have $l=1$. So the subset $(x, y, z, l=1)$, corresponds to a 4D Einstein-Hilbert theory. With the above phase space variables and using Eq.(\ref{rp1}), we can express the Friedmann equation on the brane as the constraint below:
\begin{equation}\label{cons}
x^2+y^2+z^2=1.
\end{equation}
In respect of this constraint and with attention to Eq.(\ref{nv}), we conclude that the new variables satisfy some other constraints, such that $0\leq x \leq1$, $0\leq y \leq1$ and $-1\leq z \leq1$, while $l\geq1$.

On the other hand, the Raychaudhury equation and the total equation of state (EoS) parameter of the universe, can be obtained and written in terms of the new variables as below:
\begin{eqnarray}
\frac{\dot H}{H^2}&=& -\frac{3l^2}{l^2+1}(1+z^2-y^2) \label{ray} \\
w_{tot}&=&z^2-y^2 \label{eos}
\end{eqnarray}

To build an autonomous system of ordinary differential equations, we differentiate the phase space variables in Eq.(\ref{nv}). Also, we reduce the number of degrees of freedom of the model by one, using the Friedmann constraint:
\begin{eqnarray}
  y' &=& \sqrt{\frac{3}{2}}yzl\lambda+\frac{3}{2}y(1+z^2-y^2), \label{yprime}\\
  z' &=& -3z-\sqrt{\frac{3}{2}}y^2l\lambda+\frac{3}{2}z(1+z^2-y^2), \label{zprime}\\
  l' &=&  \frac{3}{2}l\left(\frac{l^2-1}{l^2+1}\right)(1+z^2-y^2), \label{lprime}\\
  \lambda'&=&\sqrt{6}lz\lambda^2(\Gamma-1) \label{lambdaprime}
\end{eqnarray}
In these equations, the prime means derivative with respect to $\ln a$ ($a$ is the scale factor), and $\Gamma=VV_{\phi\phi}/V_\phi^2$, in which $V_{\phi\phi}$, demonstrates the second derivative of the potential with respect to the scalar field. This 4D autonomous system of equations represents the evolution of the DGP model with a quintessence scalar field, indirectly.

According to the linear stability analysis, we first solve the equations $y'=z'=l'=\lambda'=0$, simultaneously to determine the critical points of the system of equations above and their respective eigenvalues. Then, we study the behavior of the system near the critical points to describe the various kinds of possible trajectories in the phase space. Obviously, an important factor at this stage is the form of the quintessence potential, because the new variable $\lambda$, depends on the $V(\phi)$. From now on, we divide the article into two different segments: a constant $\lambda$, and a varying $\lambda$. But, first we review the case $\lambda=constant$, as the authors have investigated in \cite{Quiros2}, because our discussions for a varying $\lambda$ situation, is strongly dependent on it.

\subsection{The case $\lambda=constant$}\label{sec31}

With attention to Eq.(\ref{lambdaprime}), $\lambda=constant$, could be associated with $\Gamma=1$, as well as the special situation $\lambda=0$. The former relates to an exponential potential while the latter results in a constant quintessence potential. TABLE \ref{table:1}, shows the critical points of the model, the related eigenvalues and their existence condition, for $\lambda=constant$. We have to note that only those critical points that satisfy the constraints on the phase space variables, in addition to the Friedmann constraint have been mentioned in this table. It is clear that the critical points $CP_1$, $CP_2$ and $CP_3$, exist for all values of $\lambda$, while the existence of the critical points $CP_5$, $CP_6$ and $CP_7$, depends on the value of $\lambda$, and $CP_4$, only exists for $\lambda=0$. There is also a critical line $CL_1$, which involves $CP_4$, and exists only for $\lambda=0$, as well. By critical line we mean an infinite number of critical points with $y=1$, $z=0$, but with $l\geq1$. Also, it is obvious that all the critical points in TABLE \ref{table:1}, are the 4D solutions because of $l=1$, and it is just the critical line $CL_1$, which shows the effect of the extra dimension. Moreover, one can check that $CP_5$ coincides with $CP_1$, $CP_2$, $CP_4$, $CP_6$ and $CP_7$ when $\lambda=-\sqrt6$, $\lambda=\sqrt6$, $\lambda=0$, $\lambda=\sqrt3$ and $\lambda=-\sqrt3$, respectively.

\begin{table}[h]
 \caption{The critical points of the model for $\lambda=constant$}
  \centering{
   \begin{tabular}{|c|c|c|c|}
    \hline\hline
    Critical Points & $(y,z,l)$ & Eigenvalues & Existence\\ \hline\hline
    $CP_1$ & $(0,1,1)$ & $(3,3+\sqrt{3/2}\lambda,3)$ & any $\lambda$\\ \hline
    $CP_2$ & $(0,-1,1)$ & $(3,3-\sqrt{3/2}\lambda,3)$ & any $\lambda$\\ \hline
    $CP_3$ & $(0,0,1)$ & $(3/2,-3/2,3/2)$ & any  $\lambda$  \\ \hline
    $CP_4$ & $(1,0,1)$ & $(-3,-3,0)$ & $\lambda=0$ \\ \hline
    $CP_5$ & $(\sqrt{1-\lambda^2/6},-\sqrt6\lambda/6,1)$ & $(\lambda^2-3,\lambda^2/2-3,\lambda^2/2)$ & $-\sqrt6\leq\lambda\leq\sqrt6$ \\ \hline
    $CP_6$ & $(\sqrt6/2\lambda,-\sqrt6/2\lambda,1)$ & $(-\frac{3}{4}+3\sqrt{24-7\lambda^2}/4\lambda,-\frac{3}{4}-3\sqrt{24-7\lambda^2}/4\lambda,\frac{3}{2})$ & $\lambda\geq\sqrt{3}$ \\ \hline
    $CP_7$ & $(-\sqrt6/2\lambda,-\sqrt6/2\lambda,1)$ & $(-\frac{3}{4}+3\sqrt{24-7\lambda^2}/4\lambda,-\frac{3}{4}-3\sqrt{24-7\lambda^2}/4\lambda,\frac{3}{2})$ & $\lambda\leq-\sqrt{3}$ \\
    \hline\hline
    $CL_1$ & $(1,0,[1,\infty])$ & $(-3,-3,0)$ & $\lambda=0$\\
    \hline\hline
  \end{tabular}}
 \label{table:1} 
\end{table}

\begin{figure}[h]
\centering
\includegraphics[width=6cm]{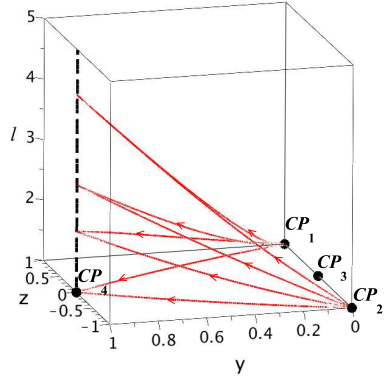}
\caption{The critical points of our dynamical system and a few trajectories for $\lambda=0$. The black dashed line represents the critical line $CL_1$.}\label{fig0}
\end{figure}

\begin{figure}[h]
\centering
\includegraphics[width=5cm]{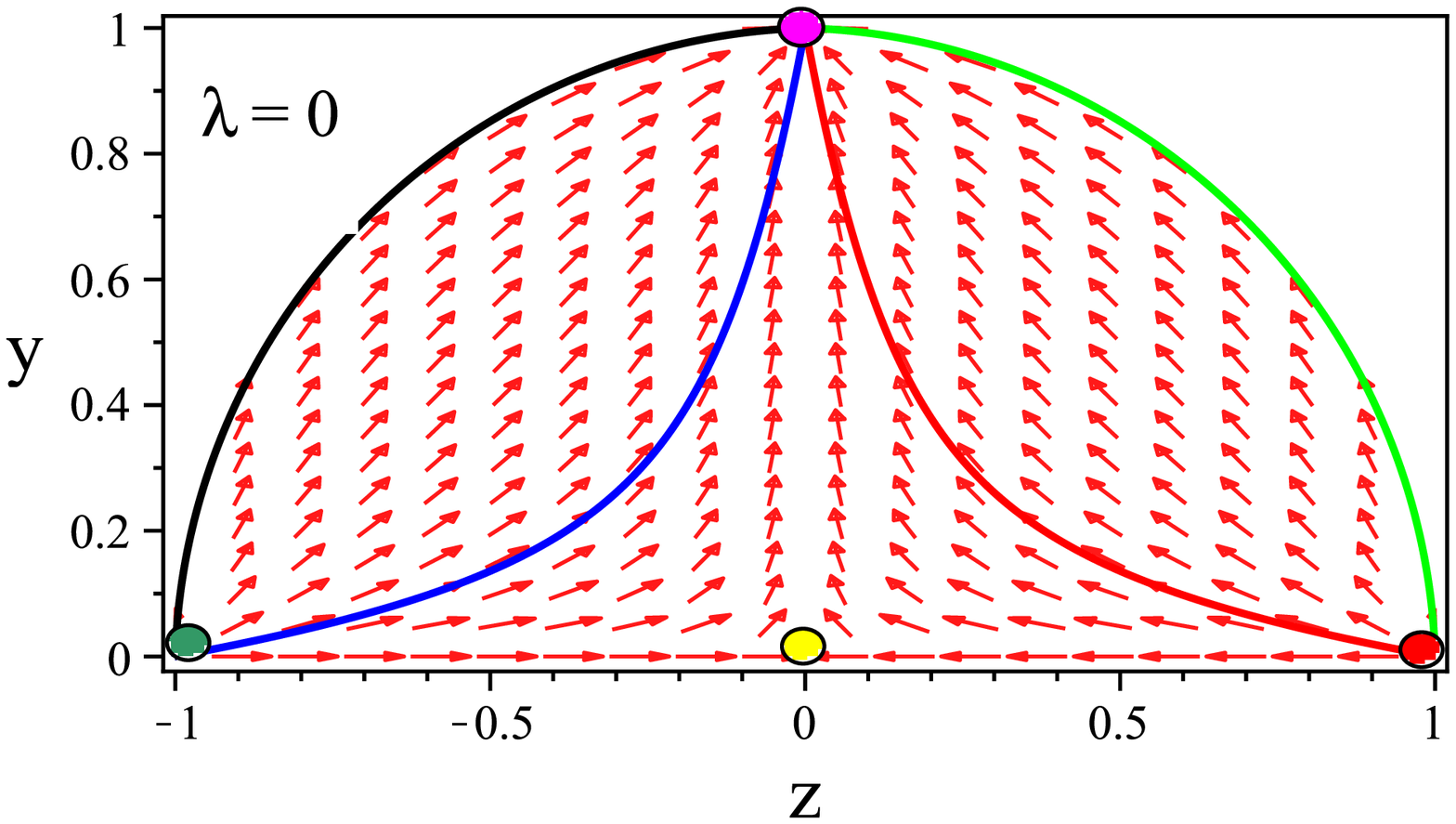}
\includegraphics[width=5cm]{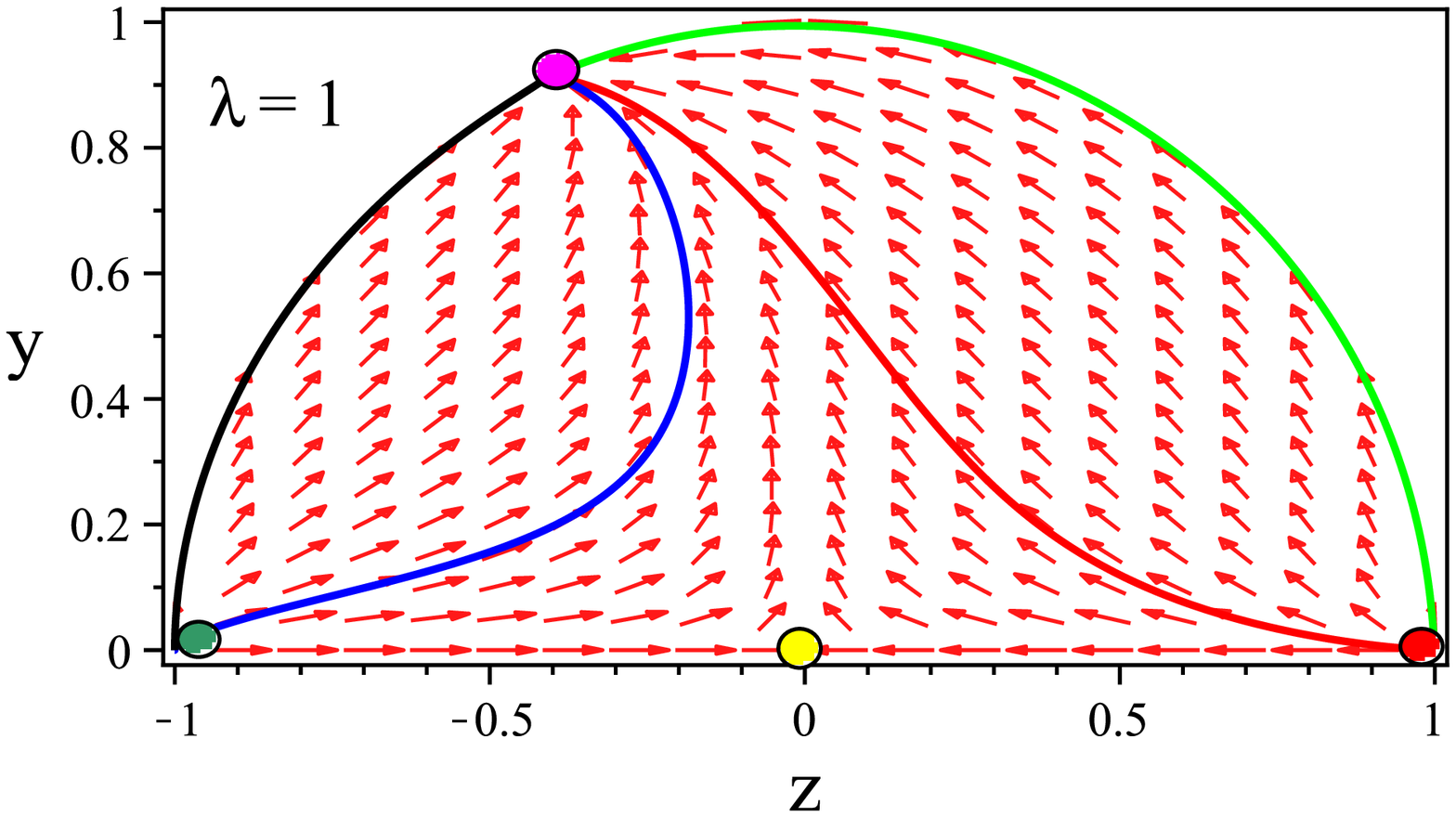}
\includegraphics[width=5cm]{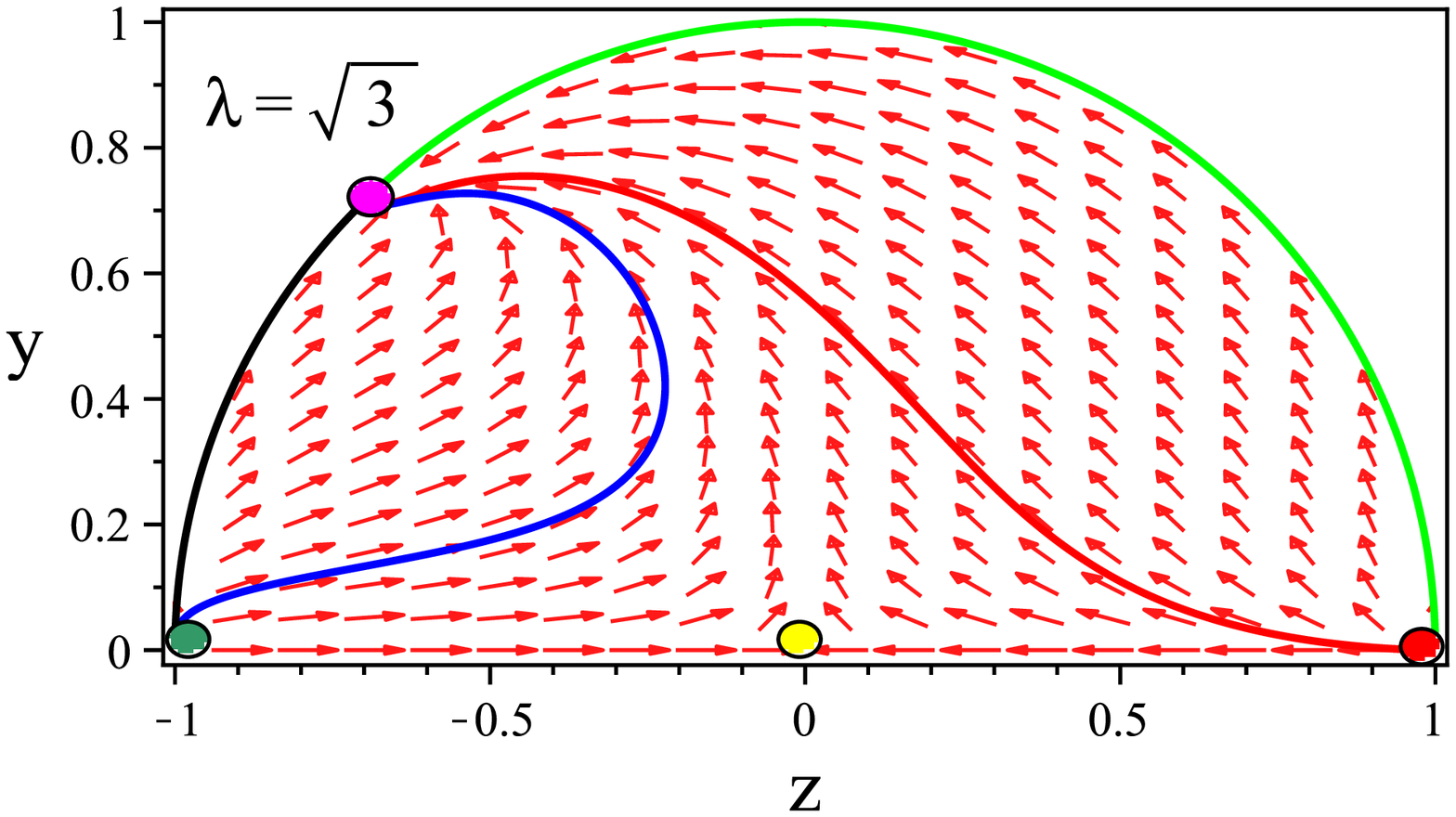}
\includegraphics[width=5cm]{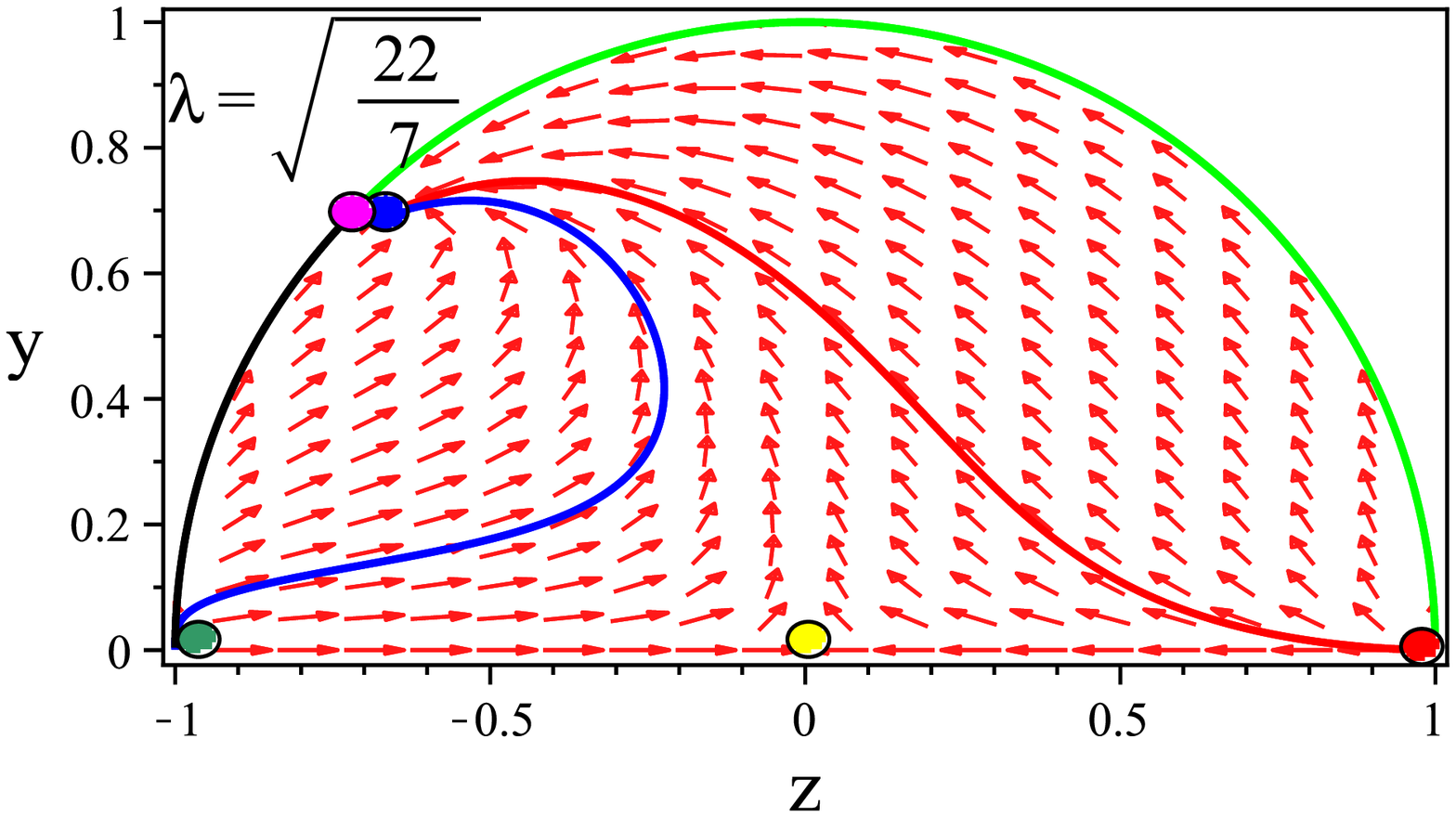}
\includegraphics[width=5cm]{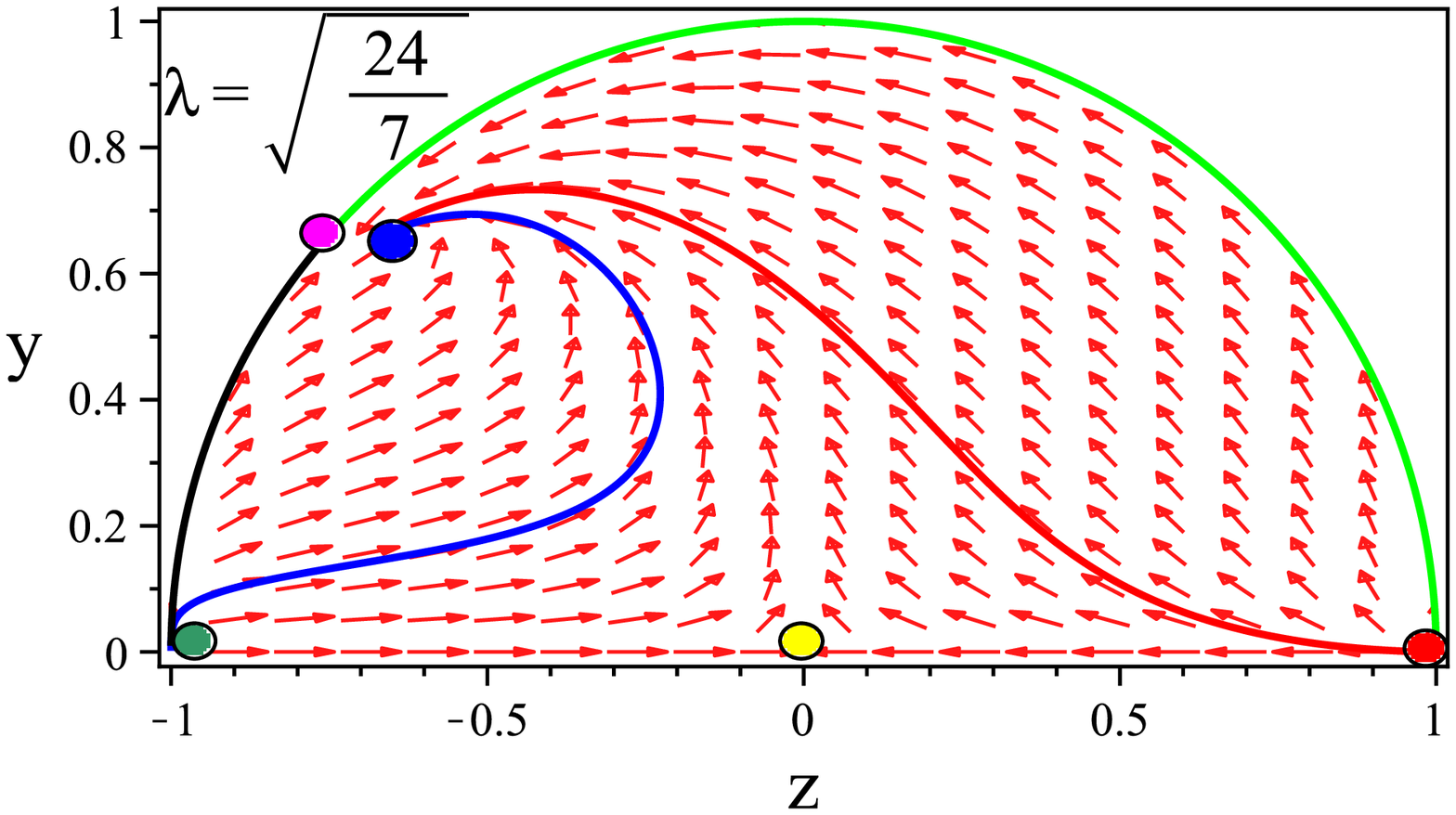}
\includegraphics[width=5cm]{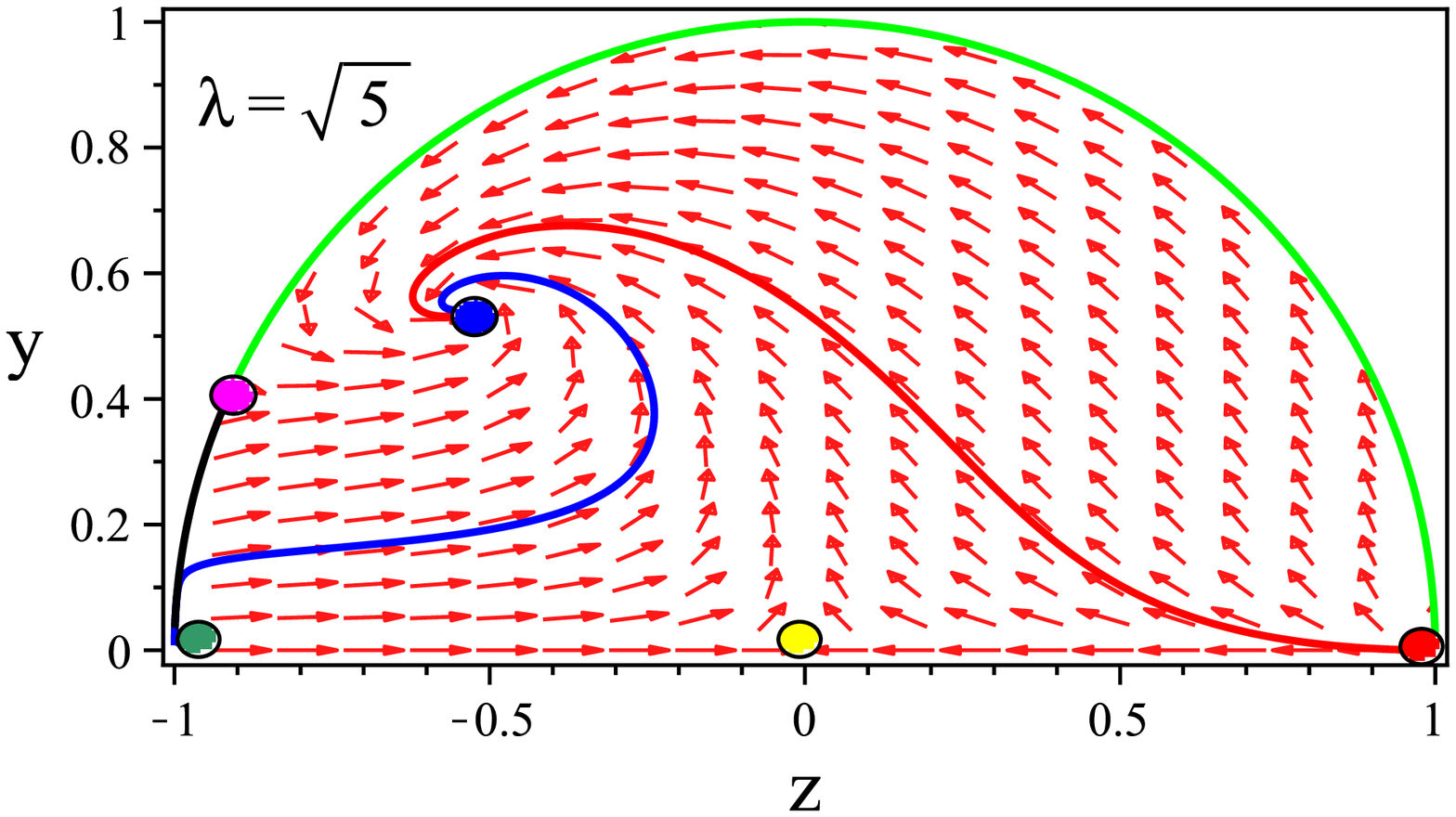}
\includegraphics[width=5cm]{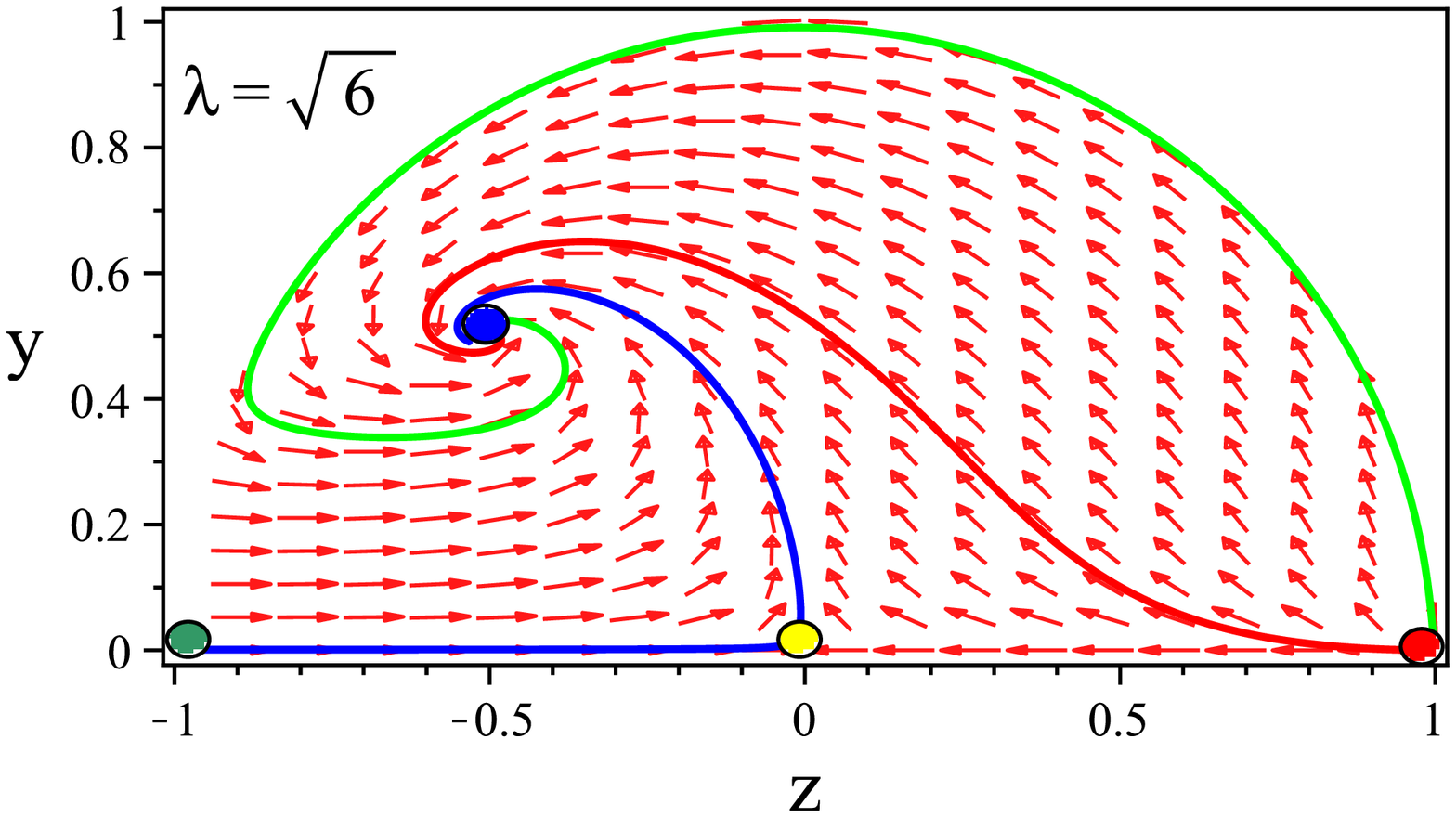}
\includegraphics[width=5cm]{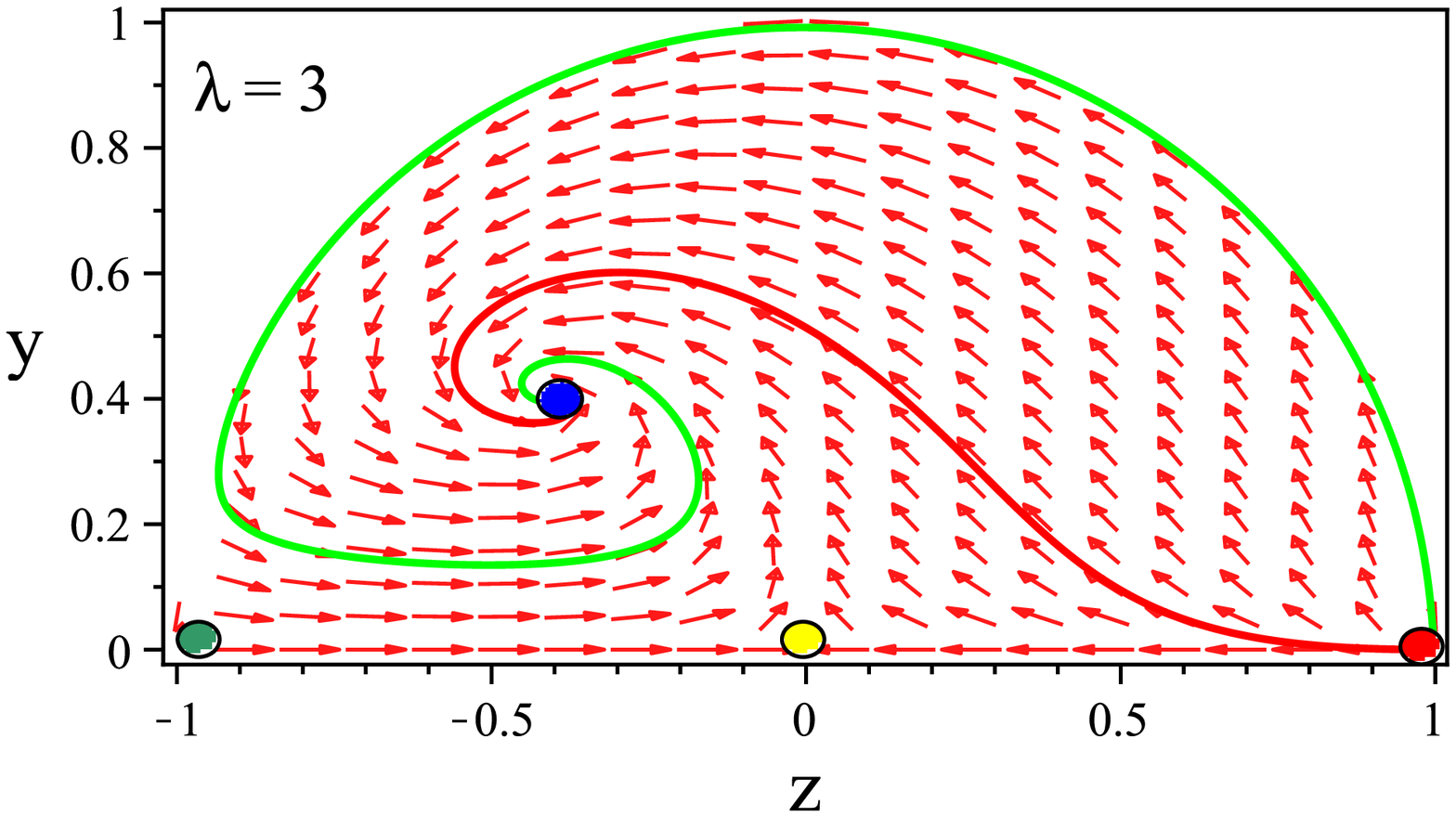}
\includegraphics[width=5cm]{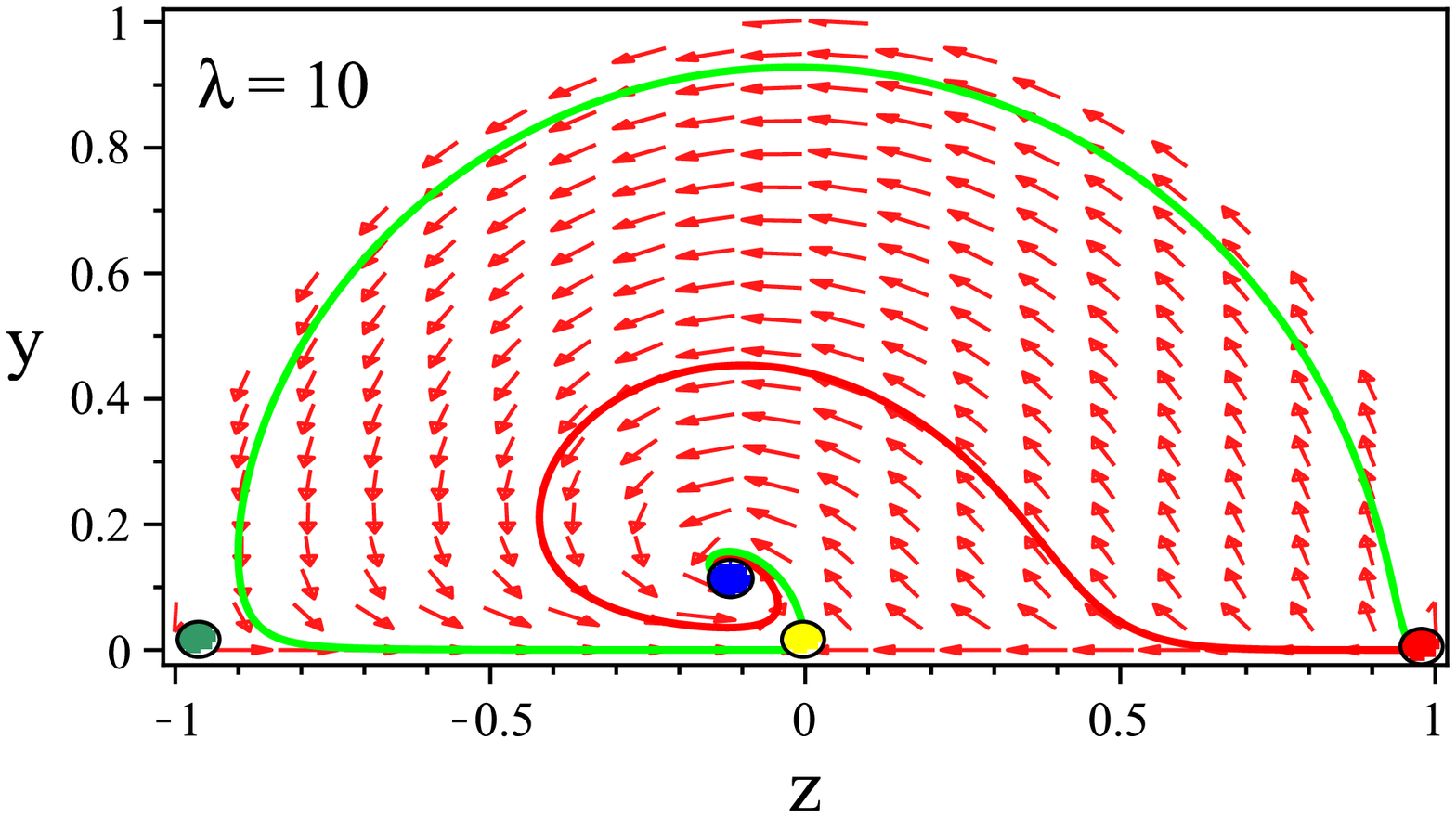}
\caption{The 2D representation of the phase space for various values of $\lambda$. The critical points $CP_1$, $CP_2$, $CP_3$, $CP_5$ and $CP_6$, have been demonstrated with a red, green, yellow, pink and blue circles, respectively. For $\lambda=0$, $\lambda=\sqrt3$ and $\lambda=\sqrt6$, the critical point $CP_5$, coincides with $CP_4$, $CP_6$ and $CP_2$, severally.}\label{fig1}
\end{figure}

The stability status of the critical points, their physical descriptions and $w_{tot}$, have been represented in TABLE \ref{table:2}. With attention to Eqs.(\ref{nv}) and (\ref{cons}), critical points $CP_1$ and $CP_2$, are kinetic dominated solutions. Also, since in these cases $w_{tot}=1$, they behave as stiff matter. Given the value of parameter $\lambda$, they may be unstable if all their eigenvalues are positive, or they could be saddle if their eigenvalues have different signs. $CP_3$ is another critical point of our model that with attention to $w_{tot}=0$, represents a matter dominated universe and is always a saddle point. We call it a pure matter dominated universe, because the Friedmann constraint for $y=0$ and $z=0$, yields $x=1$. $CP_4$ demonstrates a quintessence potential dominated solution and with attention to $w_{tot}=-1$, we can consider it as a DE dominated solution. But, in this situation we cannot specify the stability status using the common linear perturbation method, because one of its eigenvalues is zero. In such cases, one must adopt other stability approaches. Here, we resort to a numerical analysis. FIG.\ref{fig0}, illustrates some trajectories related to various initial conditions in our phase space. It is clear that all the trajectories start from the critical points $CP_1$ and $CP_2$, which are repeller for $\lambda=0$. Also, it is clear that $CP_4$, is an attractor. The critical point $CP_5$, is generally a saddle point, except for $\lambda=\pm\sqrt6$, and $\lambda=0$, because in these cases it coincides with the critical points $CP_{2,1}$ and $CP_4$. Substituting this critical point into the Friedmann constraint we obtain $x=0$, which in turn implies that $CP_5$, corresponds to a scalar field dominated solution, and satisfies the relation $y^2+z^2=1$. It means that $CP_5$, lies on a unit circle in $yz$-plane (See FIG.\ref{fig1}). Although for $CP_6$ and $CP_7$, we have $w_{tot}=0$, these critical points do not show a pure matter dominated era, because they do not result in $x=1$. We call them scaling solutions. The larger the value of $|\lambda|$, the closer to a pure matter dominated era. It is worthy to note that $CP_6$ and $CP_7$, behaves as the scalar field dominated solution $CP_5$, for $\lambda=\sqrt3$ and $\lambda=-\sqrt3$, respectively. They are always saddle critical points, though for a given range of $\lambda$, in which their eigenvalues are not real, they show a spiral behavior. The critical subset $CL_1$, which like the critical point $CP_4$ only exists for $\lambda=0$, has a zero eigenvalue, as well. Again, regarding FIG.\ref{fig0}, one can conclude that $CL_1$, is an attractor line. It is a potential dominated solution which includes the effect of the extra dimension, additionally. These results are similar to the results of \cite{Quiros2}.

\begin{table}[h]
 \caption{The stability status of the critical points}
  \centering{
   \begin{tabular}{|c|c|c|c|}
    \hline\hline
    Critical Points & $w_{tot}$ & Stability in 3D & Description\\ \hline\hline
    $CP_1$ & $1$ & {\begin{tabular} {@{}c@{}} unstable for $\lambda\geq-\sqrt6$ \\ saddle for $\lambda<-\sqrt6$ \end{tabular}} & kinetic dominated\\ \hline
    $CP_2$ & $1$ & {\begin{tabular} {@{}c@{}} unstable for $\lambda\leq\sqrt6$ \\ saddle for $\lambda>\sqrt6$ \end{tabular}} & kinetic dominated\\ \hline
    $CP_3$ & 0 & saddle & pure matter dominated \\ \hline
    $CP_4$ & $-1$ & stable & DE dominated \\ \hline
    $CP_5$ & $\lambda^2/3-1$ & saddle & scalar field dominated \\ \hline
    $CP_6$ & $0$ & {\begin{tabular} {@{}c@{}} saddle for $\sqrt{3}\leq\lambda\leq\sqrt{24/7}$ \\ spiral saddle for $\lambda>\sqrt{24/7}$ \end{tabular}} & scaling solution \\ \hline
    $CP_7$ & $0$ & {\begin{tabular} {@{}c@{}} saddle for $-\sqrt{24/7}\leq\lambda\leq-\sqrt{3}$ \\ spiral saddle for $\lambda<-\sqrt{24/7}$ \end{tabular}} & scaling solution \\
    \hline\hline
    $CL_1$ & $-1$ & stable & DE dominated\\
    \hline\hline
  \end{tabular}}
 \label{table:2} 
\end{table}

The two dimensional (2D) phase space diagrams of our model for different positive values of parameter $\lambda$, in $yz$-plane ($l=1$), have been shown in FIG.\ref{fig1}. We have ignored the negative values because of the symmetry. There is an important point that we want to explain here. As one can see in TABLE \ref{table:2}, the title of the third column is \textit{Stability in 3D}, that means all of the expressions in this column have not been written for a 2D phase plane, but rather for a three dimensional (3D) phase volume. This, in turn means that a given critical point in TABLE \ref{table:2}, may have different stability properties in a given phase plane, such as $yz$-plane. This is the case in our model, but only for those critical points that depend on $\lambda$, and this feature plays a crucial role in the evolution of the universe in our model. As we mentioned earlier, $CP_5$, is unstable for $\lambda=\pm\sqrt6$, stable for $\lambda=0$, and saddle for other values of $\lambda$, but in 3D. One can check that in 2D $yz$-plane, $CP_5$, will be a stable critical point for $-\sqrt3\leq\lambda\leq\sqrt3$, as it is clear from FIG.\ref{fig1}. Also, $CP_6$ and $CP_7$, are not saddle critical points in 2D $yz$-plane, but rather stable solutions. For instance, $CP_6$, will be a spiral stable critical point for $\lambda>\sqrt{24/7}$, and a stable critical point for $\sqrt3\leq\lambda\leq\sqrt{24/7}$, as it can be seen in FIG.\ref{fig1}. One can easily check that $CP_7$, is spiral stable for $\lambda<-\sqrt{24/7}$, and stable for $-\sqrt{24/7}\leq\lambda\leq-\sqrt{3}$.

\begin{itemize}
  \item \textit{The critical points at infinity}
\end{itemize}

\begin{figure}[h]
\centering
\includegraphics[width=6cm]{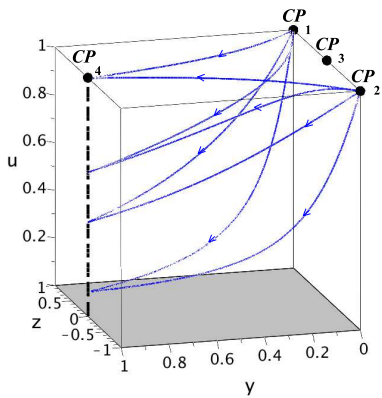}
\caption{The critical points of our dynamical system and a few trajectories for $\lambda=0$, in the new phase space. The black dashed line and the gray plane represents the critical line $CL_1$ and the critical plane $CPN$, respectively. $CPN$, results from the analysis at infinity.}\label{fig00}
\end{figure}

\begin{figure}[h]
\centering
\includegraphics[width=6cm]{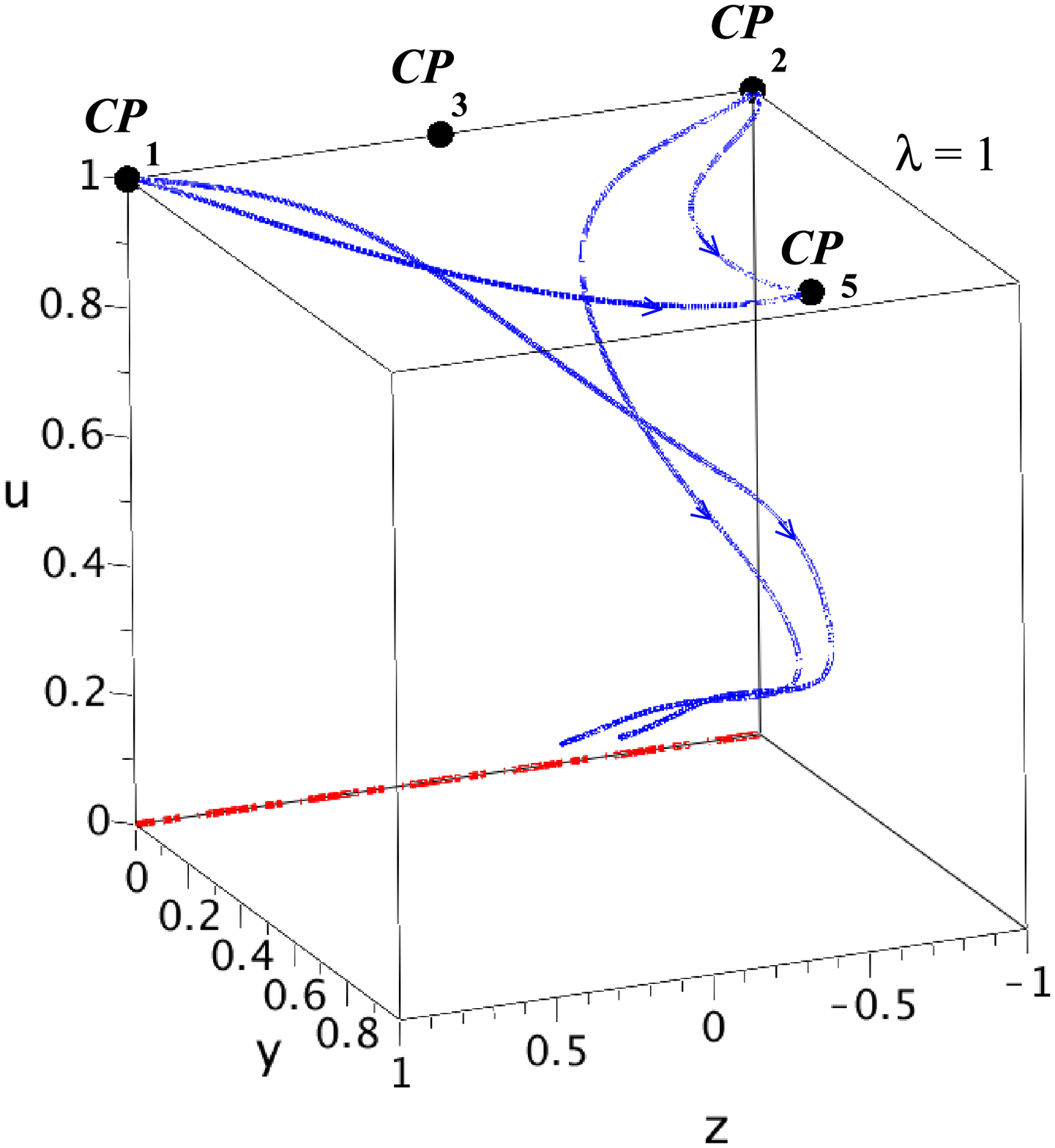}
\includegraphics[width=6cm]{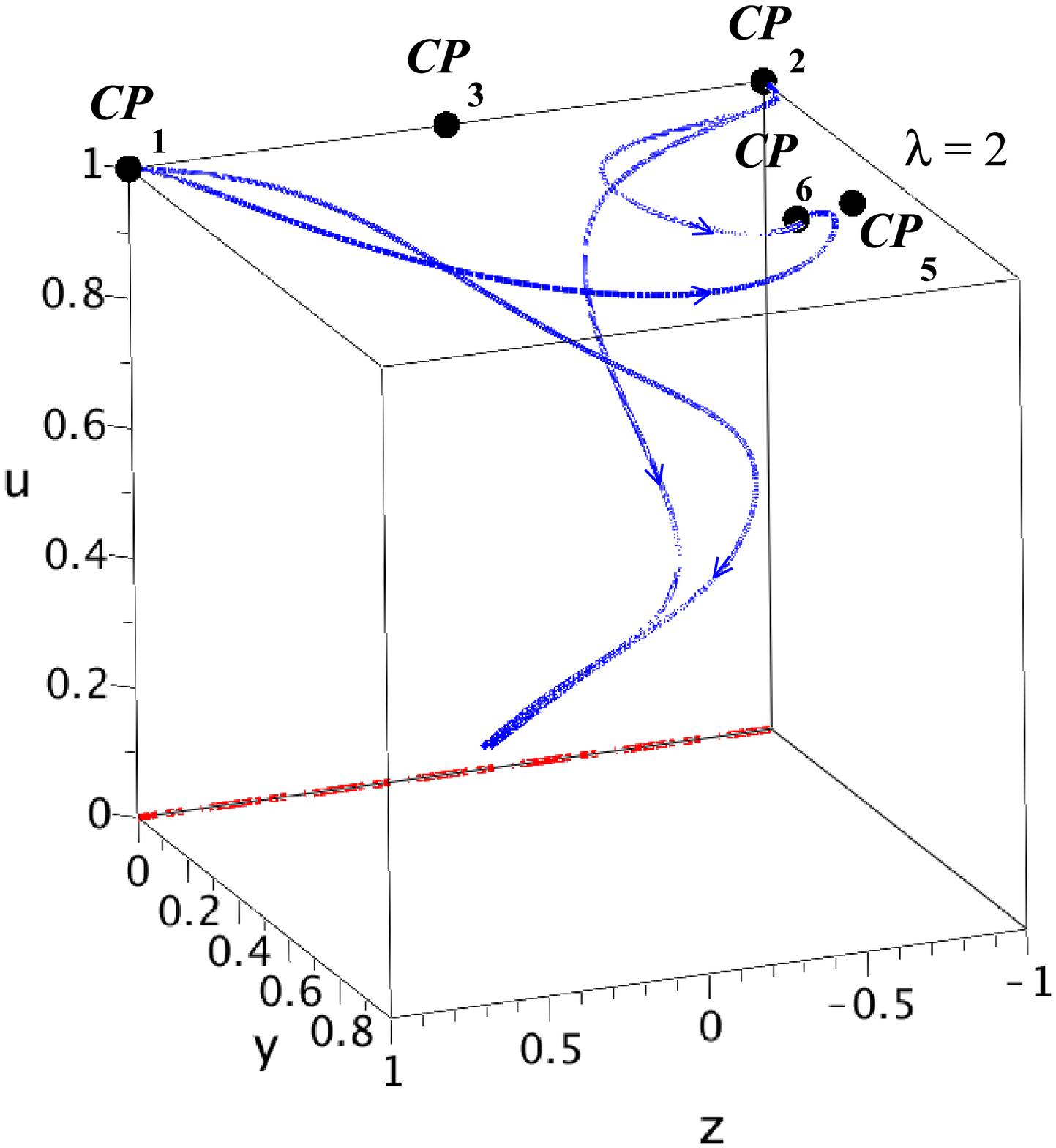}
\includegraphics[width=6cm]{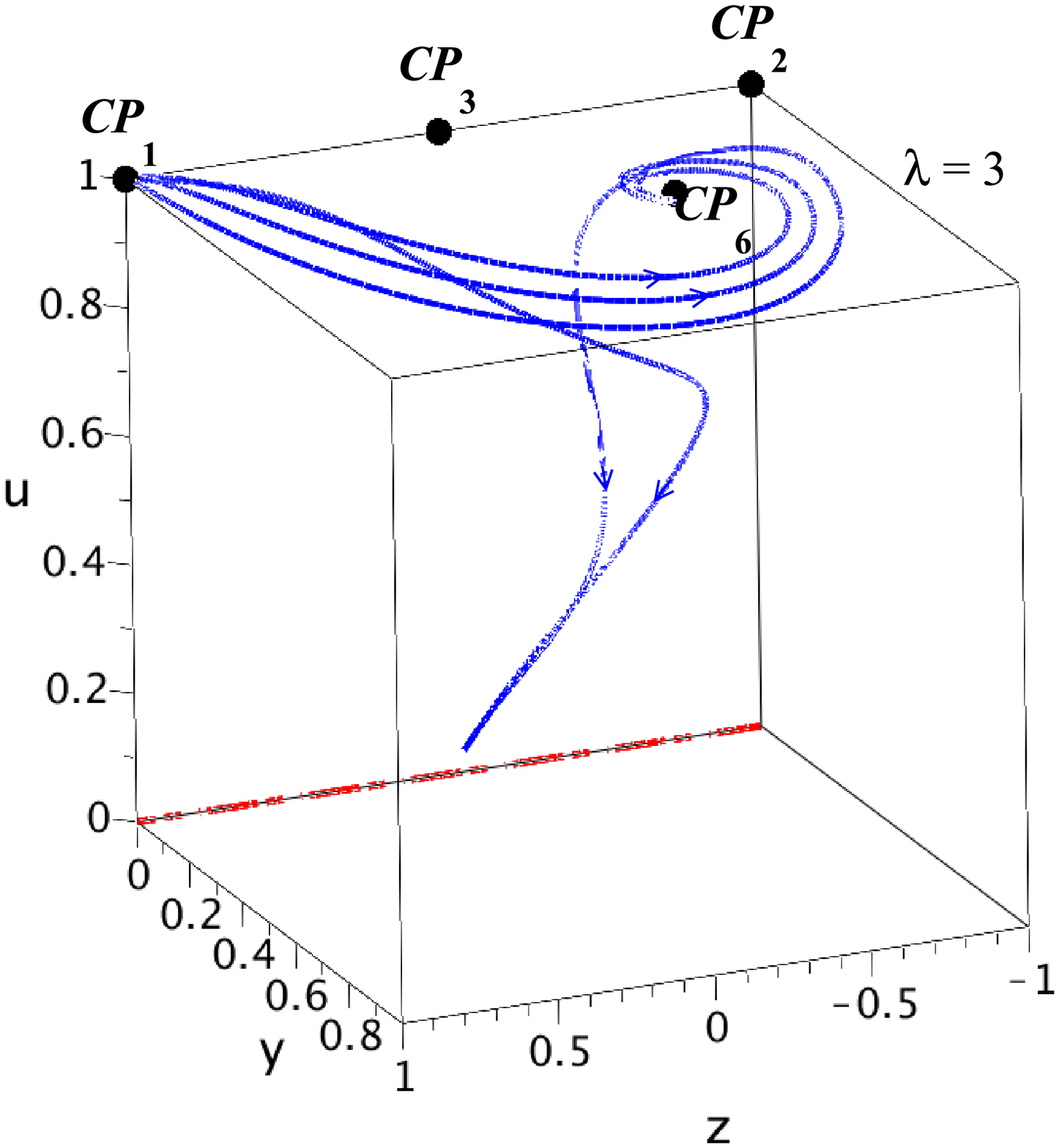}
\caption{The critical points of our dynamical system and a few trajectories for $\lambda=1,2,3$, in the new phase space. The red dash-dotted line represents the critical line $CL_2$ which results from the analysis at infinity.}\label{fig000}
\end{figure}

Since the new variable $l$ is unbounded, our discussion above is incomplete yet and we have to analyze the stability of the system at infinity, as well. We have seen the effect of the extra dimension in our model through the critical line $CL_1$ which exists only for $\lambda=0$, but what about other values of $\lambda$? The answer may be related to the analysis at infinity. Therefore, we try to compact our dynamical system defining a new variable as
\begin{equation}\label{w}
    u = \frac{1}{l}
\end{equation}
so that for $l=1$, and $l\rightarrow\infty$, we have $u=1$ and $u=0$, respectively, while $u$ satisfies the constraint $0\leq u\leq1$. Then, we obtain a new set of ordinary differential equations as below:
\begin{eqnarray}
  \frac{dy}{d\xi} &=& \sqrt{\frac{3}{2}}yz\lambda+\frac{3}{2}yu(1+z^2-y^2), \label{yprimenew}\\
  \frac{dz}{d\xi} &=& -3zu-\sqrt{\frac{3}{2}}y^2\lambda+\frac{3}{2}zu(1+z^2-y^2), \label{zprimenew}\\
  \frac{du}{d\xi} &=&  -\frac{3}{2}u^2\left(\frac{1-u^2}{1+u^2}\right)(1+z^2-y^2), \label{lprimenew}\\
  \frac{d\lambda}{d\xi}&=&\sqrt{6}z\lambda^2(\Gamma-1) \label{lambdaprimenew}
\end{eqnarray}
in which $\frac{d}{d\xi}=u\frac{d}{d\ln a}$. When we calculate the critical points of this new system we find one additional critical line for any value of $\lambda$ as $(CL_2: u=0, y=0, z=z)$, and also one critical plane for only $\lambda=0$ as $(CPN: u=0, y=y, z=z)$, on top of all the results in TABLE.\ref{table:1}. Obviously, $CL_2$, is a part of $CPN$, for $\lambda=0$. Using the Friedmann constraint we can conclude that they are matter scaling solutions, because both the quintessence scalar field and the matter content contribute in these solutions. The only difference between them is that for $CL_2$, the contribution of the quintessence potential is zero. Also, one can check that for both of them $0\leq w_{tot}\leq1$, and as a result they cannot certainly relate to an accelerated expanding phase. Since at least one of their eigenvalues is zero, we utilize the numerical approach to understand their stability characteristics. FIG.\ref{fig00}, illustrates that for $\lambda=0$, the critical plane $CPN$, and also the critical line $CL_2$, behave as saddle critical subsets. But the case differs for other values of $\lambda$. In these situations as one can see in FIG.\ref{fig000}, the critical line $CL_2$, behaves as an attractor.

Now, we have completed our analysis for the case $\lambda=constant$. We have understood that in our model the universe always starts from the unstable kinetic dominated critical points $CP_1$ and $CP_2$. But its fate depends on the value of $\lambda$, and also the initial conditions. For $\lambda=0$, it even reaches the stable DE dominated critical point $CP_4$ if it evolves in 4D, or it comes to the stable DE dominated critical line $CL_1$ which shows the effect of the extra dimension if it evolves in 5D. Also, for other values of $\lambda$, and in 5D, the universe finally approaches the matter scaling stable critical line $CL_2$, which cannot describe the current accelerated expansion. But if the universe evolves in 4D, it even reaches a scalar field dominated stable critical point $CP_5$, or it comes to a matter scaling stable critical point $CP_6$ (or $CP_7$, for negative values of $\lambda$). Depending on the value of $\lambda$, it may show the late time acceleration. This case is important in our following discussions and will be studied in detail.

\subsection{The case $\lambda=\lambda(\phi)$}\label{sec32}

If one considers the quintessence potential anything, except the constant or exponential potential, $\lambda$, will be a dynamical quantity. Here, we are interested in studying the behavior of the model in the case of a Gaussian potential. Thus we assume all the critical points in TABLE \ref{table:1}, and the ones we obtained at infinity, as the instantaneous critical points of the respective dynamical system \cite{Ravanpak2}-\cite{Ng}. With this assumption, it is clear that $CP_5$, $CP_6$ and $CP_7$, are moving critical points that indicate where the solution tends to at each instant if it evolves in 4D. Also, it is worthy to note that in the case of a Gaussian potential, $CP_4$, $CL_1$ and $CPN$, correspond to the extremum of the potential where $\lambda=0$. To understand the evolution of our universe in a varying $\lambda$ situation completely, we need to find the asymptotic behavior of $\lambda$. In other words, it is important to know that either $\lambda\rightarrow\infty$, or it approaches zero. Various kinds of potential satisfy different asymptotic limits. For potentials of the form $V=V_0\phi^{-n}$, with $n>0$, $V_\phi$, approaches zero faster than the potential itself, then $\lambda\rightarrow0$. This is the case has been investigated for instance in \cite{Macorra}, but for $n=1$. Also, a double exponential potential as $V=V_0\exp(-\alpha e^{\phi})$, as an example of the case $\lambda\rightarrow\infty$, has been studied in \cite{Macorra}, as well. Another kind of potential for which $\lambda$, goes to infinity asymptotically, that is the case of interest for us here, is the Gaussian potential, $V(\phi)=V_0\exp(-\alpha\phi^2)$, in which $V_0$ and $\alpha$, are positive constants \cite{Macorra}. For such a potential, the quintessence scalar field can roll down plus (minus) infinity with $\dot\phi>0$ ($\dot\phi<0$). Also, one can check that in this situation $\lambda=-2\alpha M_p\phi$, which yields $\lambda\rightarrow-\infty$ ($\lambda\rightarrow\infty$), at the limit $\phi\rightarrow\infty$ ($\phi\rightarrow-\infty$). Furthermore, for a Gaussian potential one can calculate $\Gamma=1-1/(2\alpha\phi^2)$, and therefore, $\Gamma-1$, is always negative. So, with attention to Eq.(\ref{lambdaprime}), the sign of $\lambda'$, depends on the sign of $z$, which in turn is proportional to $\dot\phi$. Thus we see that for both $z\gtrless0$, we have $|\lambda|\rightarrow\infty$. In the following we will only discuss the positive values of $\lambda$, because of the symmetry.

\begin{itemize}
  \item \textit{The asymptotic behavior $\lambda\rightarrow\infty$}
\end{itemize}

Regarding the shape of the Gaussian potential and since $\lambda$ is increasing one can assume that it starts from the top of the potential, the state in which $\lambda=0$. For the case $\lambda=0$, the universe tends to achieve either the stable DE dominated critical point $CP_4$, or the stable DE dominated critical line $CL_1$, with $w_{tot}=-1$, but since $\lambda$ has dynamics, it does not have enough time to get to them. If the universe evolves in 5D, the trajectories end up in the critical line $CL_2$. But the case is more complicated in 4D. The destination moves around in $yz$-plane. It starts from $CP_4$, on the critical line $CL_1$, and continues as a moving stable scalar field dominated solution $CP_5$. At the same time $w_{tot}$, is increasing. As far as $w_{tot}$, is smaller than $-2/3$, the universe experiences an accelerated phase \cite{Gumjudpai}. Along with increasing of $\lambda$, $w_{tot}$, grows as well. For $\lambda>1$, $w_{tot}$, will be greater than $-2/3$, that shows the universe is in a decelerated expanding phase. $CP_5$, still keeps moving till $\lambda=\sqrt3$. At this stage we will encounter two moving critical points $CP_5$, and $CP_6$, that coincide with one another and both behave as a single stable point in 2D phase plane with $w_{tot}=0$, which is still a scalar field dominated solution. But since then, along with increasing of $\lambda$, they will move separately. On the one hand, $CP_5$, moves around in $yz$-plane as a saddle point while the contribution of the quintessence kinetic term increases and of its potential term decreases, and at the same time $w_{tot}$, grows. Finally, when $\lambda=\sqrt6$, and $w_{tot}=1$, $CP_5$, coincides with $CP_2$, which is a kinetic dominated solution and behaves as stiff matter. On the other hand,  $CP_6$, also moves after the separation from $CP_5$, but as a stable scaling solution with $w_{tot}=0$, in which the contribution of the matter content is increasing while of the scalar field is diluting. At $\lambda=\sqrt{24/7}$, $CP_6$, turns to a spiral saddle in 3D, or in fact to a spiral stable critical point in $yz$-plane. As $\lambda$ increases, $CP_6$, slowly becomes close to $CP_3$, while it is still a spiral stable scaling solution. Finally, in the limit $\lambda\rightarrow\infty$, it approaches $CP_3$, which is a pure matter dominated solution with $w_{tot}=0$, while it behaves as a spiral attractor.

\subsection{Observational constraints}\label{sec32}

Irrespective of all the mathematical options and physical discussions above one may like to know the most realistic situations in this article. This fact that how our universe has evolved in the past and how it will do so in the future, depends on how fast our system reaches a neighborhood of one of the moving stable critical points in our model. As we mentioned in the introduction, a lot of cosmological observations have unveiled that the universe is currently experiencing a very rapidly accelerated expanding phase. Therefore, we can conclude that the evolution of the universe was fast enough, so that it had arrived a neighborhood of $CP_5$, when $\lambda$ had not yet reached $1$ and $w_{tot}$, was still smaller than $-2/3$. Because it is the only possible case of our model in which the universe can experience the accelerated expansion. But according to our results, this means that our universe will certainly undergo another phase transition from acceleration to deceleration in the future along with the evolution of $\lambda$. Will the future deceleration be a reality? The idea of the future deceleration has been theoretically studied in the literature. For instance, in the context of modified theories of gravity \cite{Srivastava},\cite{Paul} and in an extended Chaplygin gas cosmology \cite{Kahya}, the authors have shown the possibility of a transition from acceleration to deceleration in the future universe. Also, in \cite{Pan}, the authors have represented a future deceleration using the particle creation mechanism in nonequilibrium thermodynamics. Although there is not yet any direct observational evidence for this possible phase transition but many attempts have been made to address this issue by reconstructing the deceleration parameter $q$, from various observational data. For instance, in \cite{Shafieloo}-\cite{Biswas}, using different parametrization of the EoS parameter of DE, the authors have shown that the cosmic acceleration is slowing down. So, the possibility of a future deceleration is not far from expectation.

\begin{figure}[h]
\centering
\includegraphics[width=6cm]{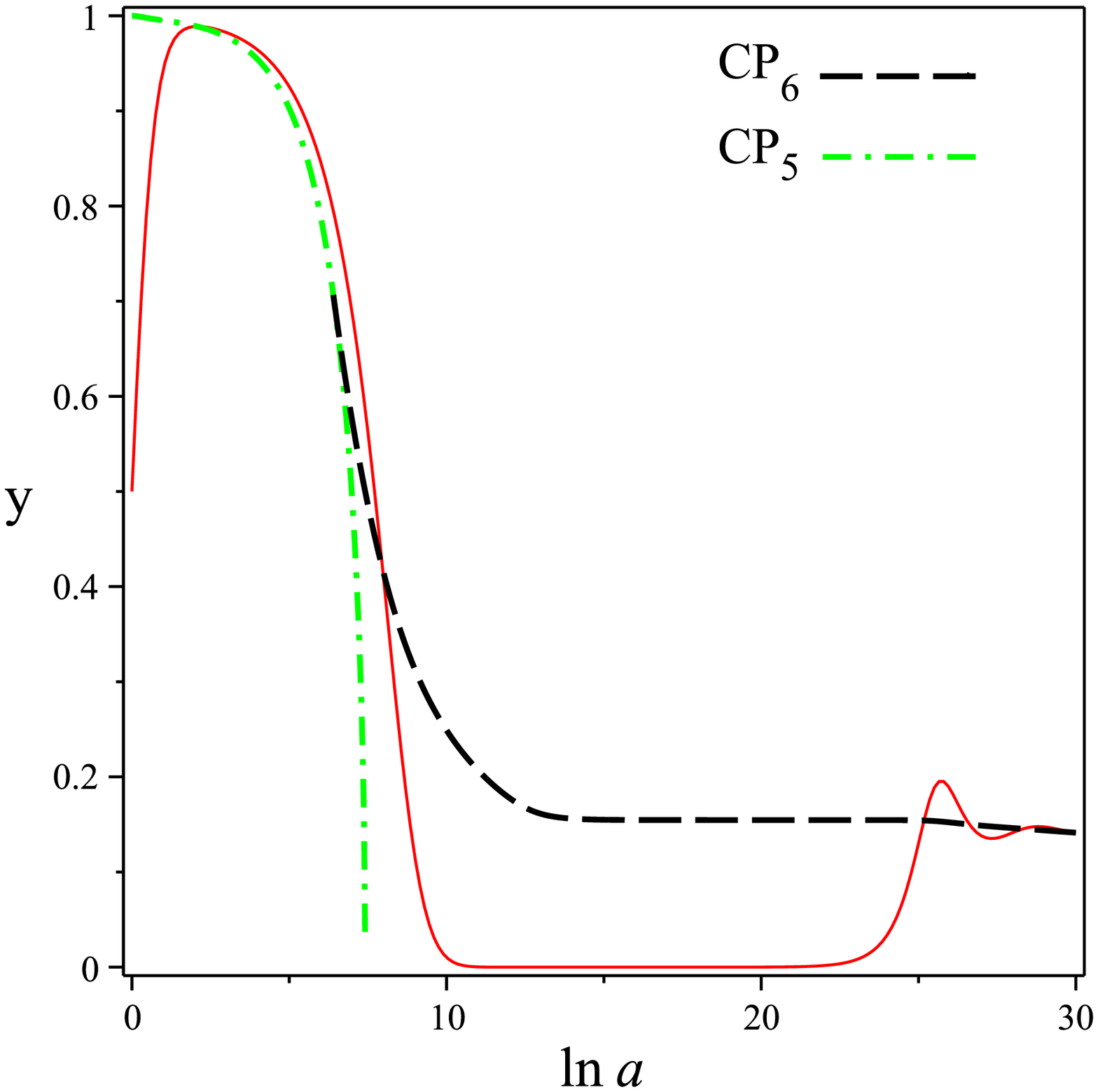}
\includegraphics[width=6cm]{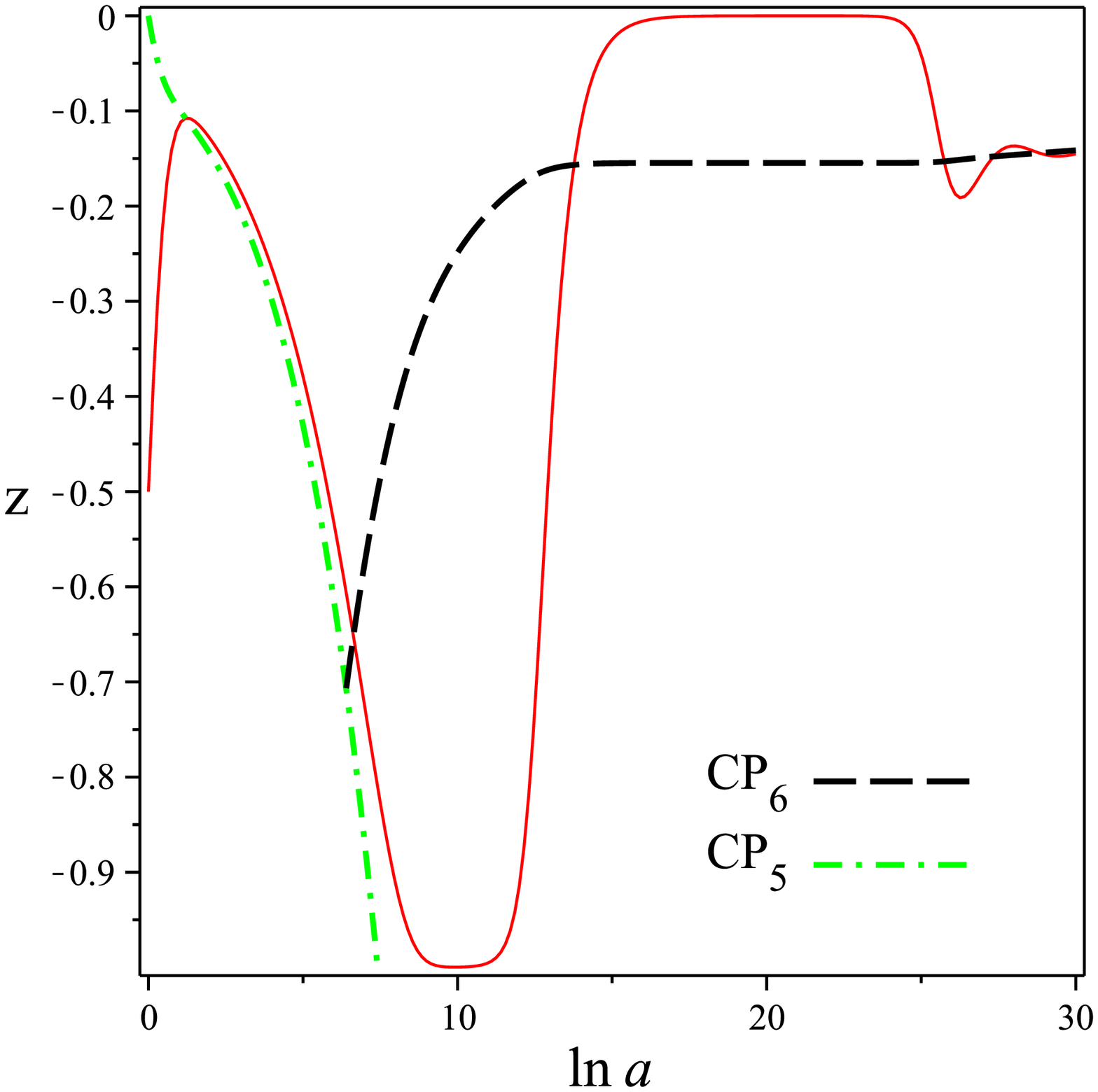}
\caption{The evolutionary curves of the dynamical variables $y$ and $z$, for the initial conditions $y=0.5$, $z=-0.5$, $\lambda=0$ and $l=1$.}\label{fig2}
\end{figure}

\begin{figure}[h]
\centering
\includegraphics[width=6cm]{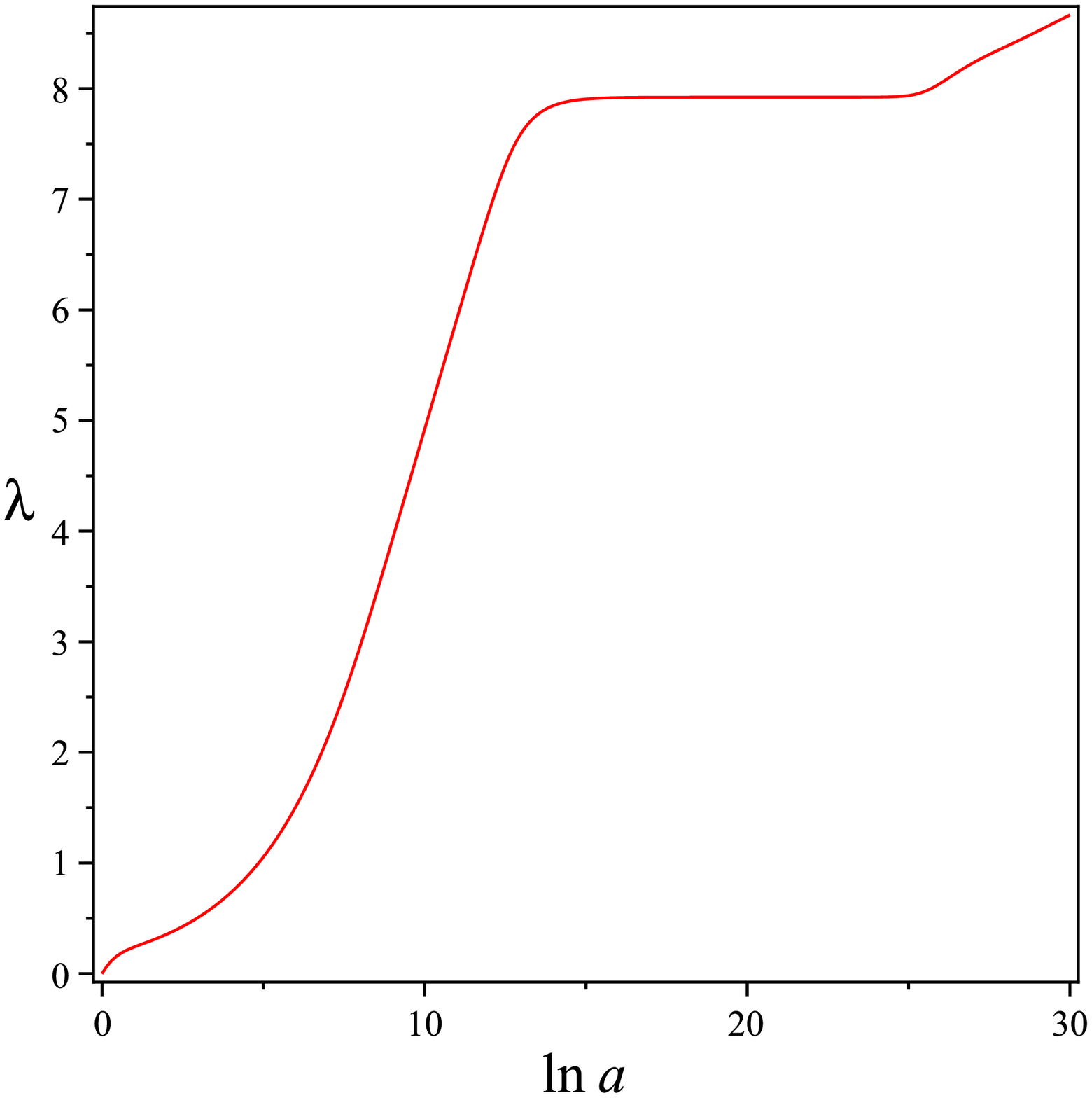}
\includegraphics[width=6cm]{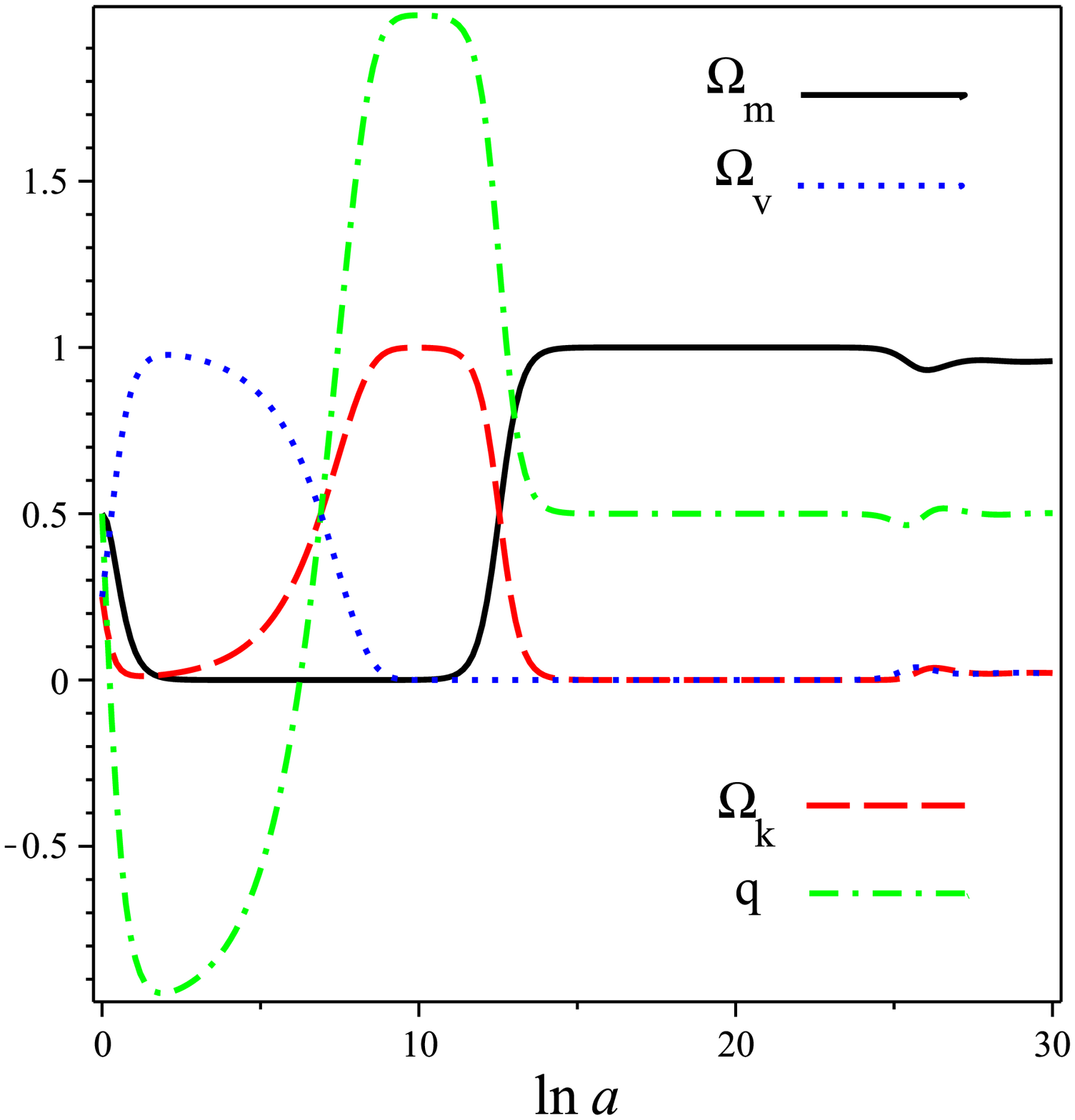}
\caption{Left: the evolutionary curve of $\lambda$. Right: the evolutionary curves of the model parameters $\Omega_m$, $\Omega_V$, $\Omega_k$ and the deceleration parameter $q$ for the initial conditions $y=0.5$, $z=-0.5$, $\lambda=0$ and $l=1$,}\label{fig3}
\end{figure}

FIG.\ref{fig2} and FIG.\ref{fig3}, illustrate the evolution of various cosmological parameters of our model for a specific choice of initial conditions. FIG.\ref{fig2}, demonstrates the evolution of two of our dynamical system variables $y$ and $z$, with respect to $\ln a$, in addition to the behavior of two moving critical points $CP_5$ and $CP_6$. What is important is that both $y$ and $z$, arrive at $CP_5$, so quickly that $\lambda$, has not yet equalled one (See FIG.\ref{fig3}), and as a consequence the universe experiences a phase of accelerated expansion. Then, they both follow the curve of $CP_5$, for a given time period. As soon as $CP_6$ appears, they start to get away from the curve of $CP_5$, and turn to the one of $CP_6$. Finally, they both catch the curve of $CP_6$ that is approaching $CP_3$.

FIG.\ref{fig3} left, shows how $\lambda$ changes with respect to $\ln a$. We see that it is always increasing though the rate of increase is not uniform and varies from one place to another. FIG.\ref{fig3} right, illustrates the evolution of our model parameters versus $\ln a$. In the beginning, the contribution of the quintessence potential ($\Omega_V=V/3M_p^2H^2$), increases while the contribution of its kinetic term ($\Omega_k=\dot\phi^2/6M_p^2H^2$), and also the one of the matter content ($\Omega_m=\rho_m/3M_p^2H^2$), decreases. Therefore, the universe enters an accelerated expanding phase very quickly as it is obvious from the curve of the decelerating parameter $q$. But after a period of time, $\Omega_V$ and $\Omega_k$, exchanges their role in the evolutionary scenario. During this process, the universe undergoes another phase transition from acceleration to deceleration. As it is clear in FIG.\ref{fig3} right, $q$, crosses zero line and takes positive values. And the fate of the universe in our model, as we discussed earlier, is a matter dominated era. One can see From FIG.\ref{fig3} right that $\Omega_m$, is the dominant component of our model at late times.

\section{summary and remarks}\label{sec4}

In this article we have studied the evolution of a normal DGP cosmological model in the presence of a quintessence scalar field DE component on the brane with a Gaussian potential, in the context of the dynamical system approach. We have derived an autonomous system of ordinary differential equations in terms of some new dimensionless dynamical variables, and have obtained the critical points of the model even at infinity. We have represented that for a Gaussian potential, the parameter $\lambda=M_pV_\phi/V$, has dynamics and approaches infinity, asymptotically. So, one can consider all the critical points, the critical lines and the critical plane for the case of a constant $\lambda$, as the instantaneous solutions, so that among them $CP_4$, $CL_1$ and $CPN$, can relate to the top of the Gaussian potential and $CP_5$, $CP_6$ and $CP_7$, can move in the phase space and are called moving critical points.

We have indicated that if our universe evolves in 5D, it ends up in the attractor line $CL_2$, a matter scaling solution in which the quintessence potential has no share, with $0\leq w_{tot}\leq1$, that clearly does not show an accelerated expansion. If we consider the 4D evolution on the $yz$-plane, we find that our universe evolves such that it continuously pursues a stable critical point. We have discussed that depending on how fast our universe evolves, it may experience an accelerated expanding phase or not, but in both the cases the fate of our universe is a matter dominated epoch without acceleration. We have illustrated that if the variation of $\lambda$, is slow enough, the universe first follows the trajectory of $CP_5$, and then turns toward to the one of $CP_6$. If our universe comes to $CP_5$, before the state $\lambda=1$, it will experience the acceleration, but then certainly undergo a phase transition to a decelerated expanding era. This is probably the case in the model under consideration regarding recent observational data and the present acceleration of the universe.

\end{document}